\documentclass[preprintnumbers,nofootinbib,
               superscriptaddress,eqsecnum,showpacs]{revtex4}
\usepackage{graphicx, fancybox, mathrsfs, amsmath, amssymb}

  \newcommand{\nn}{\notag \\}
  \newcommand{\f}{\frac}
  \newcommand{\bm}{\mathbf}
  \newcommand{\bbm}[1]{\mbox{\boldmath$#1$}}
  \newcommand{\sla}{\hspace{-0.5em} /}

\begin{document}
\title{Quasifermion spectrum at finite temperature
from coupled Schwinger-Dyson equations for a fermion-boson system}

\author{Masayasu Harada}
\author{Yukio Nemoto}
\affiliation{Department of Physics, 
Nagoya University, Nagoya, Aichi 464-8602 Japan}


\begin{abstract}
We nonperturbatively investigate a fermion spectrum at finite
 temperature in a chiral invariant linear
sigma model.
Coupled Schwinger-Dyson equations for fermion and boson are developed
in the real time formalism and solved numerically.
From the coupling of a massless fermion with a massive boson,
the fermion spectrum shows a three-peak structure at some temperatures
even for the strong coupling region.
This means that the three-peak structure which was originally found in the
one-loop calculation is stable against higher order corrections even
in the strong coupling region.
\end{abstract}

\maketitle

\section{Introduction} \label{sec:intro}

One of unexpected experimental discoveries in heavy-ion collisions
at the Relativistic Heavy Ion Collider (RHIC) is that the created matter
could be close to a perfect fluid\cite{Arsene:2004fa}.
Then, considerable changes have occurred with our understanding of 
quark-gluon plasma (QGP), which is now believed to be rather strongly coupled
or strongly interacting just above the critical temperature ($T_c$) of 
the chiral and deconfinement transitions, $T\gtrsim T_c$.
Nonperturbative results from lattice QCD seem to support this view:
The lowest charmonium state survives for $T\gtrsim T_c$, which suggests
strong correlations between quarks\cite{Asakawa:2003re}.
Furthermore, very small viscosity-entropy ratio suggested from
RHIC experiments is consistent with the result of the strongly 
coupled supersymmetric Yang-Mills theory based on the analysis
with the AdS/CFT
correspondence\cite{Kovtun:2004de}.

The present paper is concerned with spectral properties of fermions
in such a strongly coupled system, because quarks are the basic
degrees freedom of the created matter in the deconfined phase
together with gluons at RHIC.
It is not even clear whether quarks manifest themselves as
well-defined quasiparticles in such a system.
So far, there exist some theoretical investigations on this
issue\cite{Schaefer:1998wd,Mannarelli:2005pz}.
Recently, the thermal mass of the quark near $T_c$
was studied in quenched lattice QCD and was fitted well with a 
two-pole ansatz\cite{Karsch:2007wc}.
The authors and Yoshimoto investigated the quasiquark spectrum with
a strong coupling gauge theory based on the Schwinger-Dyson equation (SDE)
and found that the thermal mass is of the order $T$ and
depends on the coupling little in the strong coupling region\cite{Harada:2007gg}.
In Ref.~\cite{Kitazawa:2005mp}, it is shown in the Nambu--Jona-Lasinio
model that the quark spectrum has \textit{three peaks} near $T_c$,
which result from the coupling with fluctuations of the chiral 
condensate\cite{Hatsuda:1985eb}.

The above peculiar three-peak structure of the quark spectrum is then 
found to be rather generic at finite $T$:
It is due to the coupling with massive bosonic modes through the Landau
damping, which was analyzed in detail in Yukawa models at one 
loop\cite{Kitazawa:2006zi,Kitazawa:2007ep}.
While
the three-peak structure appears most clearly for the massless fermion,
it even appears for small massive fermions\cite{Kitazawa:2007ep}.
One of the remained issues is whether it is stable against higher order
effects.
One of the purposes of this paper is to study it in a linear sigma model
by incorporating nonperturbative effects with the 
SDE\cite{Roberts:2000aa}.
It is shown in Ref.~\cite{Kitazawa:2006zi}
that for the massless fermion,
the three-peak structure appears independently of the coupling
over some range of $T$ at the one-loop order.
Because the SDE should reproduce the perturbation theory
in the weak coupling region, the three-peak structure
would remain there even in this approach.
However, it is quite nontrivial whether it still appears
in the strong coupling region where higher order effects are not
negligible.

Another purpose of this study is to formulate coupled SDE
for fermion and boson at finite $T$.
In the past approaches, the boson propagator was not solved 
self-consistently, but fixed at the 
beginning\cite{Barducci:1989eu,Kondo:1993pd, Taniguchi:1995vf,Blaschke:1997bj,
Harada:1998zq,Ikeda:2001vc,Takagi:2002vj,Harada:2007gg,Nakkagawa:2007ti}.
As far as we know, this paper gives the first numerical result for
the coupled SDE for a fermion-boson system at finite $T$.
The point in our formulation is that while the imaginary parts of
the self-energies are evaluated through loop integrals, the real parts
are evaluated from the imaginary parts with the dispersion relation which
is a one-dimensional integral, and thus the computation is much faster than
that of the four-dimensional loop integral.
This reduction of the computational cost makes it possible to solve the coupled
SDE for fermion and boson quite fast.

The content of this paper is as follows.
We employ a linear sigma model in the chiral symmetric phase to study
the fermion spectrum at finite $T$.
In Sec.~\ref{sec:sde}, the model is introduced and the SDE are 
formulated using the real time formalism.
We use the ladder approximation to construct the SDE in which
the bare vertex is used for the self-energies.
Then, the renormalization is carried out for the boson mass and some
remarks are made for the triviality of the model.
In Sec.~\ref{sec:res}, some numerical results of the fermion spectral
functions are shown, focusing on the $T$ and coupling dependences and
nonperturbative effects by comparing with the results in the one-loop
calculation.
Concluding remarks and outlook are given in Sec.~\ref{sec:conc}.

\section{Schwinger-Dyson equations at finite temperature}
\label{sec:sde}

\begin{figure}[t]
\includegraphics[width=7cm]{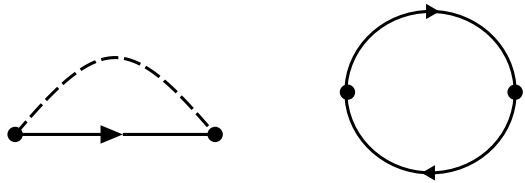}
\caption{Self-energies for fermion (left) and boson (right).
Solid lines denote the fermion propagator and dashed lines the boson
propagator.}
\label{fig:feyn}
\end{figure}

We consider a U(1)${}_L \times$ U(1)${}_R$ invariant linear sigma
model with Dirac fermions,
\begin{equation}
  \mathcal{L} = \bar{\psi} i\partial\sla \psi + \f{1}{2}\partial_\mu M^*
  \partial^\mu M -\f{m_0^2}{2}M^* M + g [\bar{\psi}_L M \psi_R +
  \bar{\psi}_R M^* \psi_L],
\end{equation}
where $M=\phi_1+i \phi_2$ and $\phi_1(\phi_2)$ denotes a scalar
(pseudoscalar) boson field.
Because our motivation lies in the quark spectrum in the QGP phase with
possible bosonic excitations, 
we assume $m_0^2>0$, i.e. the chiral symmetry is not spontaneously broken.
Self-coupling terms of the boson fields are not introduced for
simplicity. 
Their main effect is to shift the thermal mass
and the width of the boson spectrum.
As mentioned below, we renormalize the thermal mass of the boson
and measure all the dimension-full quantities in units of the renormalized
boson mass.
Then, effects of the self-coupling of the bosons are not 
important for the fermion spectrum.
Although we consider a chiral U(1)${}_L \times$ U(1)${}_R$ symmetry
for simplicity,
an extension to the SU(2)${}_L \times$ SU(2)${}_R$ symmetry is trivial
as mentioned below also.

\subsection{Formulation of the Schwinger-Dyson equations}

To construct the SDE, we employ the real time formalism.
It is convenient to use the following Green functions,
\begin{align}
  iS^>(x,y) &= \langle \psi(x) \bar{\psi}(y) \rangle, \quad
  iS^<(x,y)  = \langle \bar{\psi}(y) \psi(x) \rangle, \nn
  iG_i^>(x,y) &= \langle \phi_i(x) \phi_i(y) \rangle, \quad
  iG_i^<(x,y)  = \langle \phi_i(y) \phi_i(x) \rangle, \quad (i=1,2). 
\end{align}
Because we consider the chiral symmetric phase, there is no difference
between $G_1^\gtrless$ and $G_2^\gtrless$, and thus we simply write them as
$G^\gtrless$ in the following.

We formulate the SDE in the ladder approximation whose self-energies
are depicted in Fig.~\ref{fig:feyn}.
The fermion self-energy reads
\begin{equation}
  \Sigma(x,y) = \theta_c(x^0-y^0) \Sigma^>(x,y) 
  - \theta_c(y^0-x^0) \Sigma^<(x,y),
\end{equation}
where
\begin{equation}
  \Sigma^\gtrless(x,y) = -2g^2 i S^\gtrless(x,y) G^\gtrless(x,y),
  \label{eq:siggl}
\end{equation}
and $\theta_c$ denotes the step function along the closed time path 
in the real time formalism\cite{LeB}.
The factor 2 in Eq.~(\ref{eq:siggl}) comes from the number of boson fields,
$\phi_1$ and $\phi_2$.
Thus, for the SU(2)${}_L \times$ SU(2)${}_R$ invariant linear sigma model,
this factor is replaced by 4, which is the only difference between
the U(1)${}_L \times$ U(1)${}_R$ model and the SU(2)${}_L \times$ SU(2)${}_R$
model in the fermion self-energy.
The boson self-energy reads
\begin{equation}
  \Pi(x,y) = \theta_c(x^0-y^0) \Pi^>(x,y) + \theta_c(y^0-x^0) \Pi^<(x,y),
\end{equation}
where
\begin{equation}
  \Pi^\gtrless(x,y) = -g^2 i \textrm{Tr}[S^\gtrless(x,y) S^\lessgtr(y,x)],
  \label{eq:pigl}
\end{equation}
and Tr means the trace for the Dirac index.
For the SU(2)${}_L \times$ SU(2)${}_R$ invariant model, a factor 2 coming
from the trace in the flavor index is
multiplied in Eq.~(\ref{eq:pigl}).

In equilibrium, we can Fourier-transform the above functions using the
translation invariance.
In momentum space, the Green functions $S^\gtrless(G^\gtrless)$ have
simple expressions with the spectral function $\rho_f(\rho_b)$ and
the Fermi-Dirac (Bose-Einstein) distribution function $f(n)$,
\begin{align}
  iS^>(p^0,\bm{p}) &= [1-f(p_0)] \rho_f(p_0,\bm{p}),\quad
  iS^<(p^0,\bm{p}) = f(p^0) \rho_f(p^0,\bm{p}), \\
  iG^>(p^0,\bm{p}) &= [1+n(p_0)] \rho_b(p_0,\bm{p}),\quad
  iG^<(p^0,\bm{p}) = n(p^0) \rho_b(p^0,\bm{p}).
\end{align}
Using these expressions, the self-energies in momentum space are written as
\begin{align}
  \Sigma_\rho(P) &\equiv i\Sigma^>(P)+i\Sigma^<(P) \nn
  &= -2g^2 \int \f{d^4Q}{(2\pi)^4} [ 1-f(q^0)+n(p^0-q^0) ] 
  \rho_f(Q)\rho_b(P-Q),
  \label{eq:srho}
  \\
  \Pi_\rho(P) &\equiv i\Pi^>(P)-i\Pi^<(P) \nn
  &= -4g^2 \int \f{d^4Q}{(2\pi)^4} [ 1-f(q^0)-f(p^0-q^0) ] \rho_f^\mu(Q)
  \rho_{f\mu}(P-Q),
  \label{eq:prho}
\end{align}
with $P=(p^0,\bm{p}),\, Q=(q^0,\bm{q}),\, \rho_f^\mu=(\rho^0_f,\bbm{\rho}_f)$ 
and
$\rho_f=\gamma_0 \rho^0_f  - \bbm{\gamma}\cdot\bbm{\rho}_f$.
The functions $\Sigma_\rho$ and $\Pi_\rho$ are related to the corresponding
retarded self-energies,
\begin{align}
  \textrm{Im}\Sigma^R(P) &= -\f{1}{2}\Sigma_\rho(P) ,\quad
  \textrm{Re}\Sigma^R(P)  = \textrm{P} \int \f{dq^0}{2\pi}
  \f{\Sigma_\rho(q^0,\bm{p})}{p^0-q^0}, \nn
  \textrm{Im}\Pi^R(P) &= -\f{1}{2}\Pi_\rho(P) ,\quad
  \textrm{Re}\Pi^R(P)  = \textrm{P} \int \f{dq^0}{2\pi}
  \f{\Pi_\rho(q^0,\bm{p})}{p^0-q^0},
\end{align}
where P denotes the principal integral.

The retarded Green functions are expressed with the retarded 
self-energies and thus $\Sigma_\rho$ and $\Pi_\rho$ as
\begin{align}
  S^R(P) &= \f{1}{P\sla + \textrm{Re}\Sigma^R(P)+i\textrm{Im}\Sigma^R(P)}, \nn
  G^R(P) &= \f{1}{P^2-m_0^2+\textrm{Re}\Pi^R(P)+i\textrm{Im}\Pi^R(P)}.
\end{align}
The spectral functions are in turn expressed with the retarded Green functions,
\begin{align}
  \rho_f(P) &= -2\textrm{Im}S^R(P), \quad
  \rho_b(P)  = -2\textrm{Im}G^R(P).
  \label{eq:rhofb}
\end{align}
Equations (\ref{eq:srho})-(\ref{eq:rhofb}) are closed form for the spectral 
functions $\rho_f$ and  $\rho_b$, and thus give the self-consistency condition.

Two comments are in order.

One is that
this formulation is similar to that given in Appendix D of 
Ref.\cite{Juchem:2003bi}.
The main difference is in the calculation method of the self-energies:
While we evaluate them as momentum integrals, in Ref.\cite{Juchem:2003bi}
 the self-energies
are Fourier-transformed and evaluated in the coordinate space.
In the latter, the fast Fourier transformation algorithm can be used
for the numerical calculation and in general is faster than the 
direct evaluation of the four-dimensional integral.
In equilibrium, however, one of the angle integrals is trivial and thus
the direct evaluation of the integrals is effective as well.

The other is that in this method, only the imaginary parts of the
self-energies are evaluated through $\Sigma_\rho$ and $\Pi_\rho$.
The real parts are obtained from the imaginary parts by the
Kramers-Kronig relation which is given by the one-dimensional
integral.
Therefore, this is much efficient than the method in which both the
real and imaginary parts of
the retarded self-energies are evaluated with loop integrals.
This is the main reason why we can compute the coupled SDE
at finite $T$ with a less numerical cost.

\subsection{Renormalization} \label{sec:renorm}

In Fig.~\ref{fig:feyn}, the boson self-energy is quadratically divergent and
the fermion self-energy is logarithmically divergent, and thus
both have to be regularized.
In the numerical calculation, we introduce an ultraviolet cutoff
in the momentum integrals.
While the linear sigma model is renormalizable perturbatively,
it is known that there is no nontrivial continuum limit.
Here, we do not pursue this issue and the renormalization is carried out
 partially:
We only renormalize the quadratically divergent boson self-energy using
the once subtracted dispersion relation,
\begin{equation}
  \textrm{Re}\Pi^R(p_0,p) =
  c + (p_0^2-\alpha^2) \textrm{P} \int_0^\Lambda
  \f{dq_0}{2\pi} \f{q_0 \Pi_\rho(q_0,p)}{(p_0^2-q_0^2)(q_0^2-\alpha^2)},
\end{equation}
where $p=|\bm{p}|$ and 
we have used a property that $\Pi_\rho(p_0,\bm{p})$ is an odd function
for $p_0$.
The subtraction constant $c$ and the renormalization point $\alpha$
are determined by imposing the `on-shell' renormalization condition,
\begin{equation}
  c=0,\quad
  \alpha=\sqrt{p^2+m^2},\quad
  m=m_0,
\end{equation}
where `on-shell' means that the renormalization is done so that
the boson has the dispersion relation $p_0=\sqrt{p^2+m^2}$ 
including thermal effects.
In other words, we renormalize the boson mass such that the thermal
mass of the boson should be $m$.
This $T$-dependent renormalization is contrasted to the $T$-independent
renormalization used in the one-loop 
calculation where the $T$-independent
terms can be separated from the others and are only 
renormalized\cite{Kitazawa:2006zi,Kitazawa:2007ep}.
In the Schwinger-Dyson (SD) approach, however, 
the $T$-independent and $T$-dependent terms 
cannot be separated from each other by the self-consistency condition and
thus such a renormalization is hard to impose.
Because we are mainly interested in the fermion spectrum, we employ this simple
`on-shell' renormalization for the boson mass.

The boson self-energy is still logarithmically divergent after the
above 
renormalization and the fermion self-energy is also logarithmically divergent,
both of which are in principle renormalized with the wave function renormalization.
We, however, leave an ultraviolet three-momentum cutoff
as in the previous work\cite{Harada:2007gg}.
One reason is that numerical calculations do not suffer from such a
logarithmic divergence very much.
The other is that the wave function renormalization
modifies only the over-all factor in the propagator, which may not change the
relative peak structure of the spectral function.
Our main interest in this paper is to investigate the fermion spectrum at low
energy and low momentum where thermal effects rather than quantum effects are
dominant.

Therefore, the self-energies are written with the cutoff as follows:
\begin{align}
  \Sigma_{\rho}^0(p_0,0)
  &= -\f{g^2}{2\pi^3}\int_{-\Lambda}^\Lambda dq_0 \int_0^\Lambda dq 
  q^2 \left[1-f(q_0)+n(p_0-q_0) \right] 
  \rho_0(q_0,q) \rho_b(p_0-q_0,q), \\
  \Sigma_{\rho}^0(p_0,p) 
  &= -\f{g^2}{4\pi^3p}\int_{-\Lambda}^\Lambda dq_0 
  \int_0^\Lambda dq \int_{|p-q|}^{p+q} dk
  qk \left[1-f(q_0)+n(p_0-q_0) \right]
  \rho_0(q_0,q)\rho_b(p_0-q_0,k), \\
  \Sigma_{\rho V}(p_0,0) &= 0, \\
  \Sigma_{\rho V}(p_0,p)  
  &= -\f{g^2}{8\pi^3 p^2}\int_{-\Lambda}^\Lambda dq_0 \int_0^\Lambda dq 
  \int_{|p-q|}^{p+q} dk k (p^2+q^2-k^2)
   \left[1-f(q_0)+n(p_0-q_0) \right]
   \rho_V(q_0,q) \rho_b(p_0-q_0,k), \\
  \Pi_{\rho}(p_0,0)
  &= -\f{g^2}{\pi^3} \int_{-\Lambda}^\Lambda dq_0 \int_0^\Lambda dq 
   q^2 [1-f(q_0)-f(p_0-q_0)] [\rho_0(q_0,q) \rho_0(p_0-q_0,q)
  +\rho_V(q_0,q) \rho_V(p_0-q_0,q)], \\
  \Pi_{\rho}(p_0,p)
  &= -\f{g^2}{2\pi^3 p}\int_{-\Lambda}^\Lambda dq_0 
  \int_0^\Lambda dq \int_{|p-q|}^{p+q} dk  [1-f(q_0)-f(p_0-q_0)] \nn
  &\ \times \bigg[
  qk \rho_0(q_0,q) \rho_0(p_0-q_0,k) 
  - \f{p^2-q^2-k^2}{2} \rho_V(q_0,q) \rho_V(p_0-q_0,k) \bigg]
\end{align}
with $\Sigma_\rho(p_0,p) = \gamma_0 \Sigma^0_\rho(p_0,p) 
- \bbm{\gamma}\cdot \hat{p} \Sigma_{\rho V}(p_0,p)$,\,
$\rho_{f}(p_0,p) = \gamma^0 \rho_0(p_0,p) 
- \bbm{\gamma}\cdot \hat{p} \rho_V(p_0,p)$ and $\hat{p}=\bm{p}/p$.
It is noted that more strictly, the cutoff $\Lambda$ is introduced so that
the self-energies $\Sigma_\rho(p_0,p), \Pi_\rho(p_0,p)$ 
and the spectral functions $\rho_{0,V,b}(p_0,p)$ have nonzero
values in the range $-\Lambda\leq p_0 \leq \Lambda$ and
$0\leq p\leq \Lambda$.

As mentioned above, while the boson mass is renormalized,
the coupling constant $g$ is not and thus depends on the cutoff $\Lambda$.
In the numerical calculation given in the next section, 
we take $\Lambda/m_0\approx 10$ which is 
determined from the fact that larger values of $\Lambda$ are hard to calculate
for the numerical accuracy and smaller values of $\Lambda$ would affect thermal
effects significantly because we set $T/m_0 \lesssim 2$.%
\footnote{
As mentioned in the main text, 
the triviality means the coupling $g$ at infrared goes to zero for
$\Lambda\to\infty$.
However,
if we focus on the spectrum for $p_0, p, m \ll T$, the temperature
would be an effective ultraviolet cutoff, and thus the system would be
renormalizable even nonperturbatively
owing to effective three-dimensional reduction.
Although we consider the spectrum up to $T\approx m$, we hope that such an
effective reduction partly works.}
We have checked that the following results of the fermion spectrum
are stable under small change of the values of $\Lambda$, because we focus on
the energy-momentum region where thermal effects such as the Landau damping 
are dominant.

\section{Numerical results} \label{sec:res}

In this section, 
we examine how the quasiparticle picture of the
fermion changes by studying the fermion spectral function
as temperature or the coupling changes.
It is useful in the following analysis to employ the
expression
\begin{equation}
  \rho_{\pm}(p_0,p) = \rho_0(p_0,p)\pm\rho_V(p_0,p),
\end{equation}
where $\rho_+$ and $\rho_-$ represent the spectrum of the
fermion and the antifermion excitations, respectively,
 and are obtained from
the fermion self-energies\cite{Kitazawa:2006zi} as
\begin{equation}
  \rho_{\pm}(p_0,p) = -2\textrm{Im}\bigg[\f{1}{p_0\mp p-\Sigma_\pm(p_0,p)}
  \bigg],
\end{equation}
with 
$\Sigma_\pm=\Sigma^R_0\mp \Sigma^R_V$.
Because we have a relation $\rho_-(p_0,p)=\rho_+(-p_0,p)$ by the
charge conjugation invariance, we show only the numerical results for
$\rho_+$ below.

\subsection{Temperature dependence}

Before showing the numerical results, we briefly review the
fermion spectrum at $T=0$ and high-$T$ limit in our model.
At $T=0$, the renormalization prescription for the boson mass
given in Sec.~\ref{sec:renorm} becomes exactly the on-shell
renormalization, and thus the poles of the fermion propagator
are on the light cone.
The peaks of the fermion spectral functions $\rho_+$ and
$\rho_-$ are then
along the light cone, $p_0=p$ and $p_0=-p$, respectively.
Effects of the interaction through the SDE are contained
in the widths of the peaks and the continuum part.

In the high-$T$, weak coupling limit, the fermion spectrum approaches
that calculated in the HTL approximation with the thermal mass of the
fermion $m_T=\sqrt{2}gT/4$\cite{Weldon:1982bn}.
Both $\rho_+$ and $\rho_-$ have two $\delta$-functions corresponding to
the normal quasiparticle and plasmino excitations, in addition to the
continuum part in the spacelike region.
In the high-$T$ limit but with fixed coupling $g$, these two peaks have
widths of the order of $gm_T$\cite{LeB}.
Even in the SD approach, the fermion spectrum should approach that in the HTL
approximation for the weak coupling because the self-energies shown in
Fig.~\ref{fig:feyn} include all the hard thermal loops of this model and
the perturbation theory should 
work.

\begin{figure}[t]
\includegraphics[width=6.0cm]{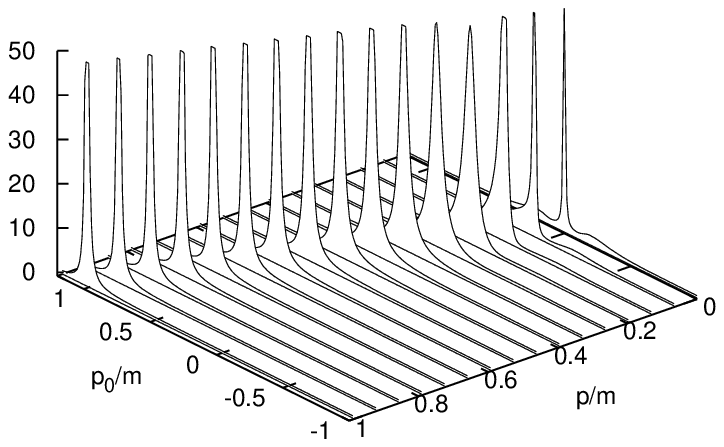} \hspace{-12mm}
\includegraphics[width=6.0cm]{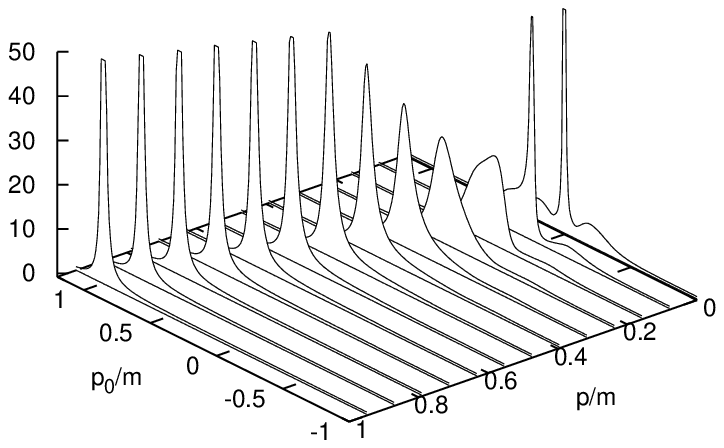} \hspace{-12mm}
\includegraphics[width=6.0cm]{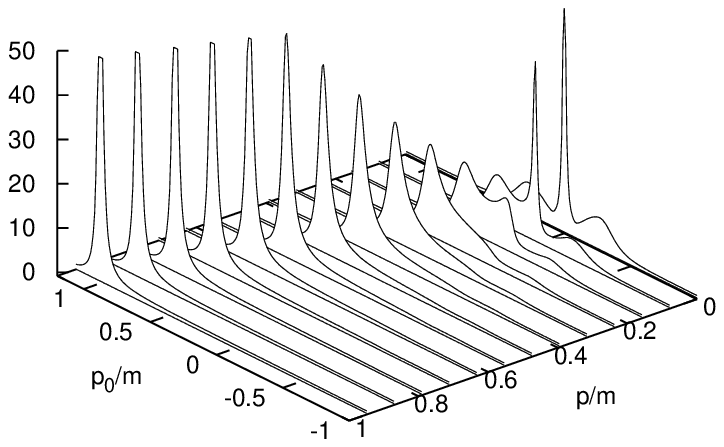}   \hspace{-12mm}
\includegraphics[width=6.0cm]{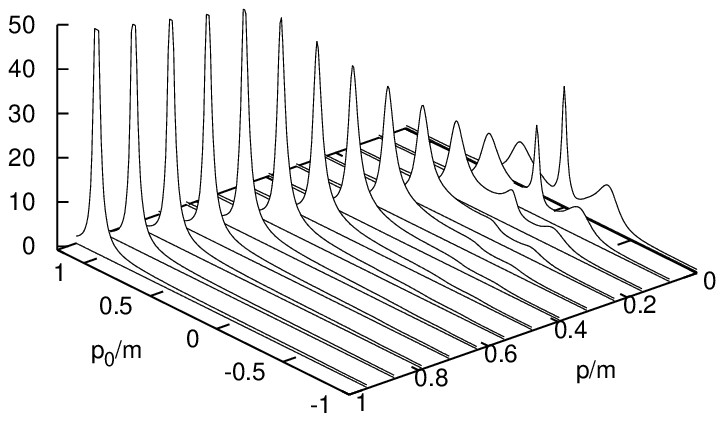} \hspace{-12mm}
\includegraphics[width=6.0cm]{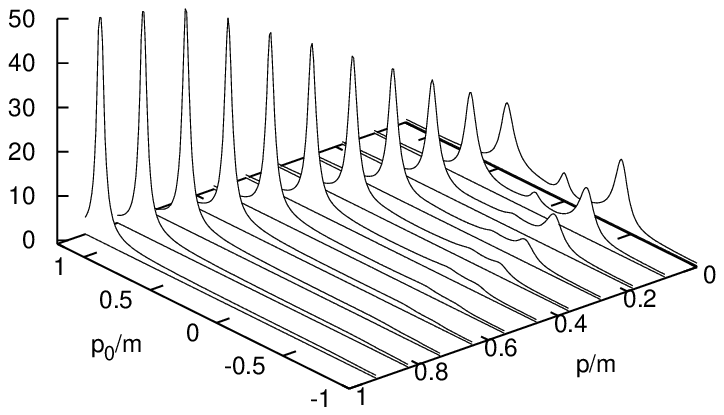} \hspace{-12mm}
\includegraphics[width=6.0cm]{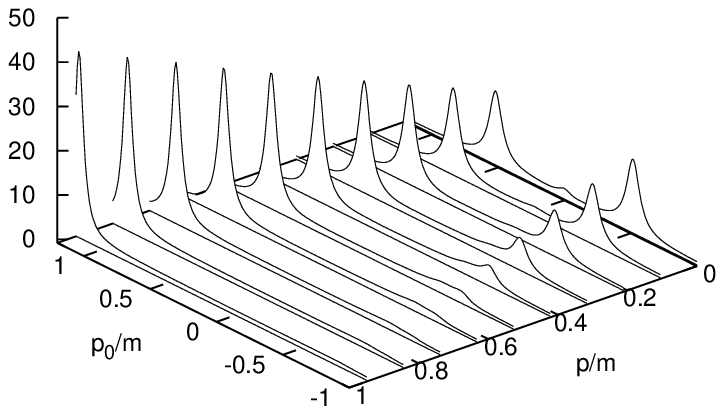}
\caption{The fermion spectral functions $\rho_+ m$ for
$g^2=1$ and $T/m=0.6$ (top left), 0.8 (top middle), 1.0 (top right),
1.2 (bottom left), 1.6 (bottom middle), 2.0 (bottom right).
The figure is clipped at $\rho_+ m=50$.}
\label{fig:tspc}
\end{figure}

\begin{figure}[ht]
\includegraphics[width=8cm]{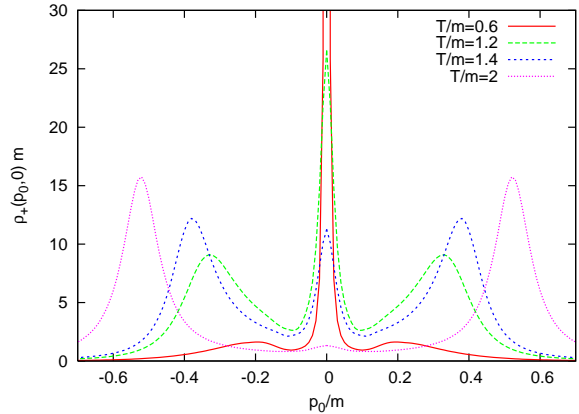}
\caption{Temperature-dependence of the fermion spectral function
$\rho_+ m$ for $p=0, g^2=1$.}
\label{fig:tspc2}
\end{figure}

\begin{figure}[ht]
\includegraphics[width=7.8cm]{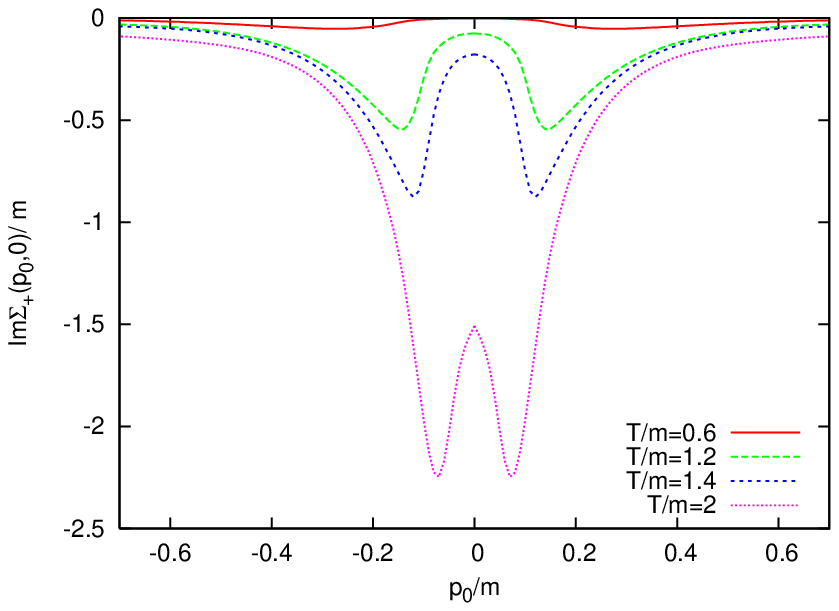}
\includegraphics[width=7.8cm]{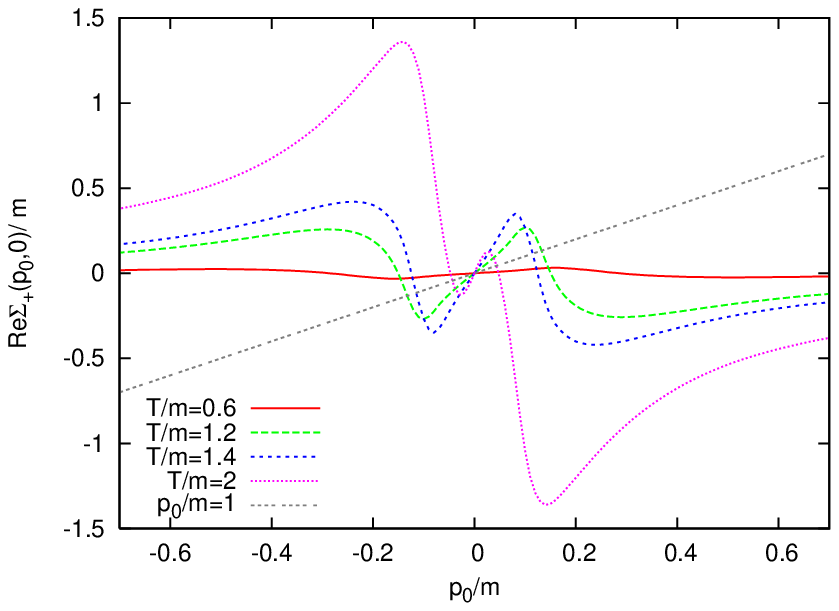}
\caption{The fermion self-energy 
$\Sigma_+/m$ for $p=0,\, g^2=1$: the imaginary part (left panel)
and the real part (right panel).}
\label{fig:sigg1}
\end{figure}

\begin{figure}[ht]
\includegraphics[width=4cm]{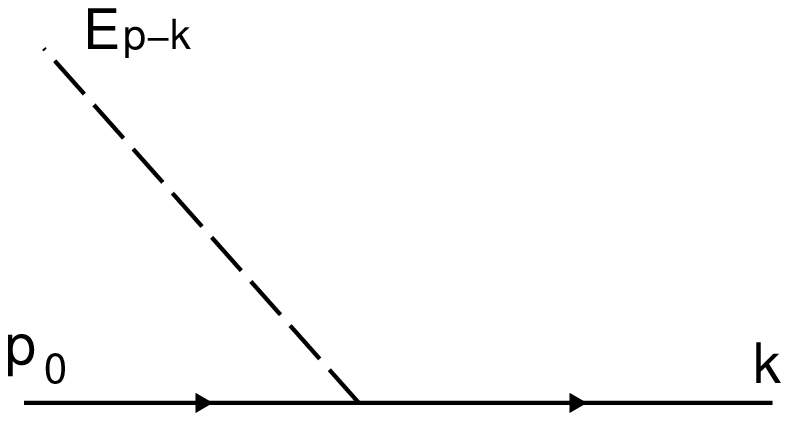}
\hspace{1cm}
\includegraphics[width=5.3cm]{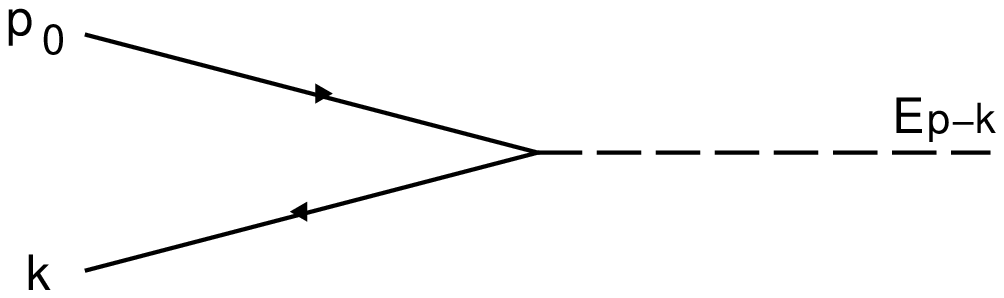}
\caption{The kinetic processes by the Landau damping 
contained in the imaginary part of the
fermion self-energy at one loop. (The corresponding inverse processes
are not shown.)
The solid line represents the fermion and the dashed one the on-shell boson
with the energy $E_{p-k}=\sqrt{(\bm{p}-\bm{k})^2+m^2}$.
The fermion with the energy $k$ is also on-shell.}
\label{fig:landau}
\end{figure}

\begin{figure}[ht]
\includegraphics[width=8cm]{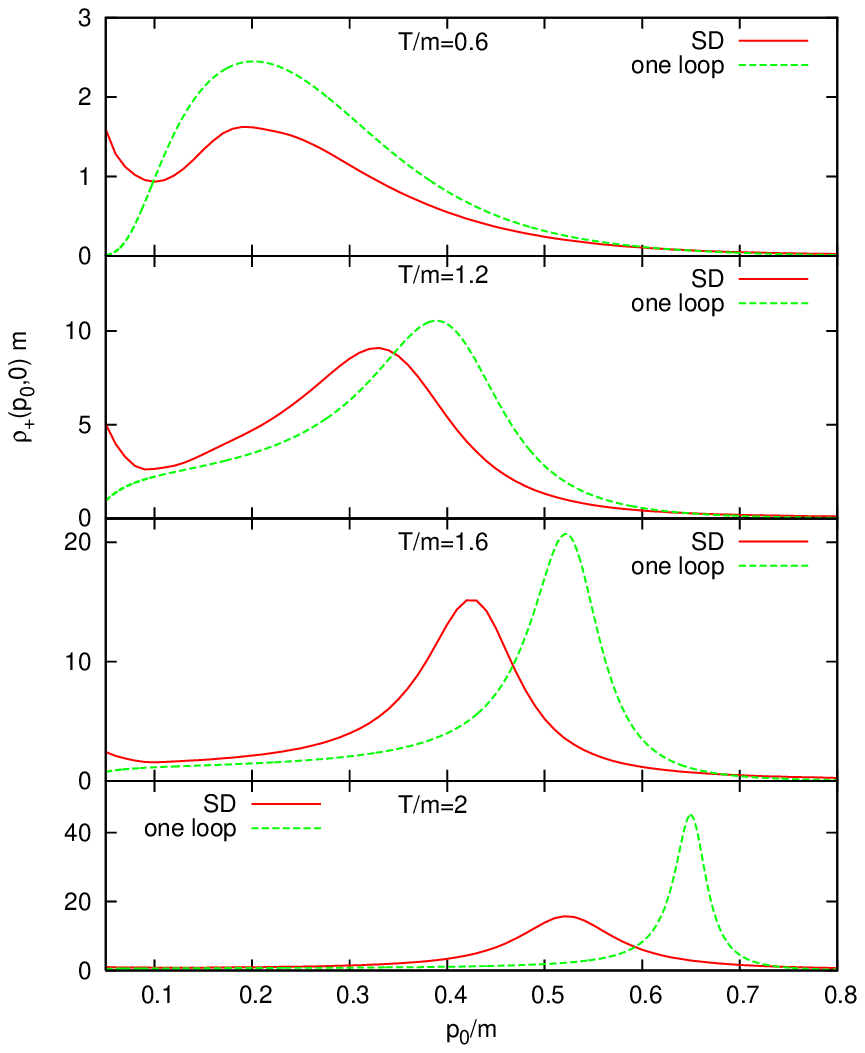}
\caption{The spectral functions $\rho_+(p_0,0) m$ for $g^2=1$ and
$T/m=0.6, 1.2, 1.6, 2$ from top to bottom.}
\label{fig:peakt}
\end{figure}

In Fig.~\ref{fig:tspc}, we plot the fermion spectral function $\rho_+$
for various temperatures.
At $T/m=0.6$, the peak position of the spectral function $\rho_+$
is almost on the light cone;
$\rho_+$ has sharp quasiparticle peaks around $p_0=p$.
At $T/m=0.8$, we see that 
the quasiparticle peaks are deformed near $p/m=0.2$
and that the peak position deviates from the light cone for
$p/m\lesssim 0.2$.
This deformation is more remarkable as $T$ is higher:
At $T/m=1.0$ and $1.2$, the quasiparticle peaks start to split 
for $p/m\lesssim 0.2$, and there appear broad peaks both in the
positive and negative energy regions around $p=0$.
Therefore, we see a \textit{three-peak structure} in the low-momentum region.
At $T/m=1.6$, three peaks clearly separate from one another and the strength
of the central peak becomes weaker than those of the peaks in both sides.
At $T/m=2$, the central peak almost disappears and thus only two peaks
remain in the low-momentum region.
These two peaks correspond to the normal quasifermion and the antiplasmino
seen in the spectrum in the HTL approximation\cite{Klimov:1981ka}.
The fact that these peaks survive for the strong coupling region is consistent
with that in Refs.\cite{Harada:2007gg, Peshier:1998dy}.

In the same figure, as momentum is higher,
the peak of the normal quasifermion becomes larger and sharper.
This is because thermal effects get weaker and thus the spectrum should
approach the free one up to the width due to quantum corrections  
as momentum is higher.
The antiplasmino peak in turn rapidly suppresses as momentum is higher, 
which is
consistent with the behavior of the HTL approximation where the residue
of the (anti-)plasmino peak suppresses exponentially with momentum.

The above result resembles that obtained at the one-loop 
order\cite{Kitazawa:2006zi}.
Thus, higher order effects do not modify the qualitative peak-structure
of the fermion spectrum at this coupling.
There is, however, a quantitative difference for the central peak:
While it is a $\delta$-function shape at the origin in the one-loop 
calculation, which can be shown analytically, it has a finite width in the
present result. 
We plot $\rho_+$ at zero momentum in Fig.~\ref{fig:tspc2} from which
we see that the width of the peak at the origin becomes broader and
the peak height smaller as $T$ increases.
It is noted that the behavior of the peak height is consistent with that
in the one-loop calculation where the residue of the $\delta$-function
peak becomes smaller as $T$ increases.
The behavior of the peak height can be found from that of the
self-energy at zero momentum shown in Fig.~\ref{fig:sigg1}:
We see that $|$Im$\Sigma_+(p_0,\bm{0})|$ at the origin increases
as $T$ increases, which means that the decay amplitude of the
fermion increases with $T$.
This is contrasted to the imaginary part of the self-energy at one loop 
which vanishes at the origin independently of $T$\cite{Kitazawa:2006zi}.
The reason of the vanishing at one loop is as follows.
We can readily show that the fermion with the energy $-m<p_0<m$ and 
the zero momentum decays through the Landau damping shown in
Fig.~\ref{fig:landau} at the one-loop order.
It is noted that the boson and the out-going fermion are on-shell in 
this figure.
Then, in the left process of Fig.~\ref{fig:landau},
 the boson momentum $k$ must be
infinity for $p_0=0$ due to the energy-momentum conservation.
Because this process and the inverse one give the statistical factor
$n(\sqrt{k^2+m^2})+f(p_0+\sqrt{k^2+m^2})$ in the amplitude,
these processes are forbidden for the infinite $k$.
Likewise, in the right process of Fig.~\ref{fig:landau},
 the momentum of the in-coming fermion
must be infinity for $p_0=0$ and then the statistical factor
vanishes as in the left process.
Therefore, we have always Im$\Sigma_+(0,\bm{0})=0$ at one loop.
Because Re$\Sigma_+(p_0,\bm{0})$ is an odd function by the charge conjugation
invariance and thus vanishes at the origin at any order,
the spectral function has a $\delta$-function shape at the origin.
In the SD approach, on the other hand, higher loop effects do not forbid 
the decay of the fermion with $p_0=p=0$ and thus Im$\Sigma_+(0,\bm{0})$ can
be finite, while Re$\Sigma_+(0,\bm{0})=0$ due to the property of the 
odd function as shown in the right panel of Fig.~\ref{fig:sigg1}.
This is the reason why there can exist a peak with a finite width at 
the origin in the SD approach.
(In the right panel of Fig.~\ref{fig:sigg1}, 
the line $p_0/m=1$ is also plotted. Crossing points
between this line and Re$\Sigma_+$ give spectral peaks if
Im$\Sigma_+/$Re$\Sigma_+$ is sufficiently small there.)
We note, however, that it is nontrivial whether the three-peak
structure appears in the SD approach because it would disappear
if the value of Im$\Sigma_+(0,\bm{0})$ is quite large even for $T/m<2$.
The result shown in Fig.~\ref{fig:tspc} indicates that the three-peak
structure is stable against higher order corrections at this coupling.

Next we examine the peaks in both sides which correspond to
the normal quasifermion for $p_0>0$ and the anti-plasmino for
$p_0<0$.
In Fig.~\ref{fig:peakt}, we plot the spectral functions $\rho_+$ for
$p_0>0$ and $p=0$ obtained from the SDE and the one-loop
calculation.
Because we have a relation $\rho_+(p_0,\bm{0})=\rho_+(-p_0,\bm{0})$,
the anti-plasmino peak is also shown in this figure by replacing $p_0$
with $-p_0$ in the horizontal axis.
At $T/m=0.6$ and $1.2$, the peak position and the width from the
SDE are similar to those from the one-loop calculation.
For higher $T$, the peak position and the width from the SDE 
are smaller and broader than those from the one loop, respectively.
The broadening of the peak from the SDE could be understood
from the fact that the probability of the boson emission and 
absorption from a fermion increases owing to multiple scatterings
with bosons through the self-consistency 
condition\cite{Harada:2007gg}.
The decrease of the peak position for higher $T$ due to nonperturbative 
effects is a new result, for which we do not have intuitive
explanation in the present.
As seen in the next subsection, this behavior of the decreasing peak
position is also seen when the coupling $g$ becomes larger.

\subsection{Coupling dependence}

\begin{figure}[t]
\includegraphics[width=6.0cm]{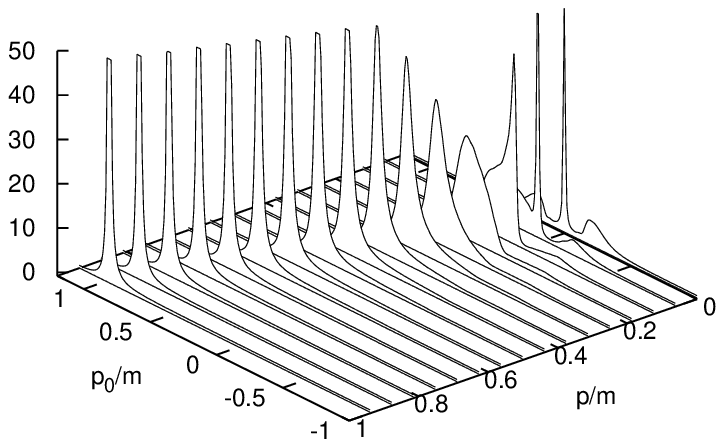} \hspace{-12mm}
\includegraphics[width=6.0cm]{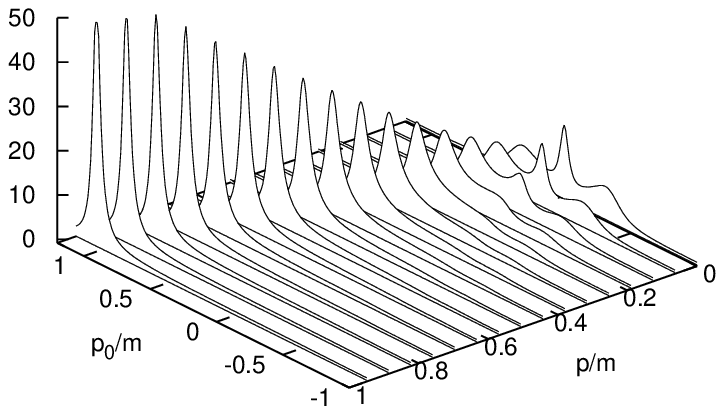} \hspace{-12mm}
\includegraphics[width=6.0cm]{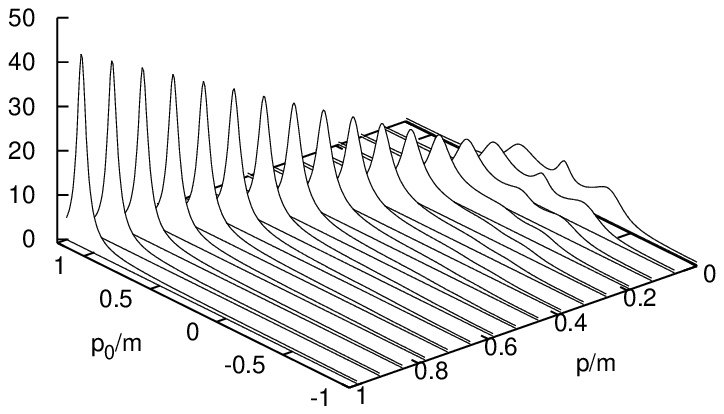}
\caption{The fermion spectral functions $\rho_+ m$ for
$T/m=1$ and $g^2=0.4$ (left), 2 (middle), 4 (right).
The figure is clipped at $\rho_+ m =50$.}
\label{fig:gspc}
\end{figure}

\begin{figure}[ht]
\includegraphics[width=7.8cm]{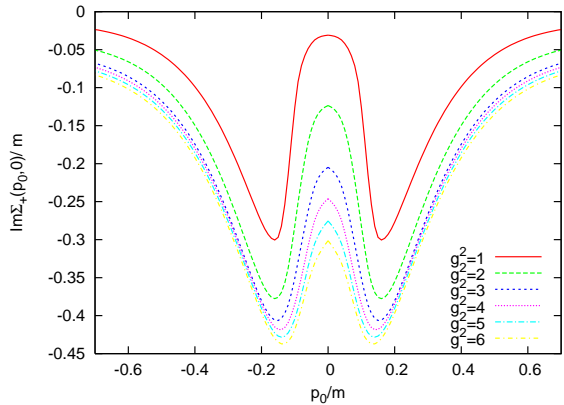}
\caption{The imaginary part of the fermion self-energy 
$\Sigma_+/m$ for $p=0,\, T/m=1$.}
\label{fig:sigg2}
\end{figure}

\begin{figure}[ht]
\includegraphics[width=8cm]{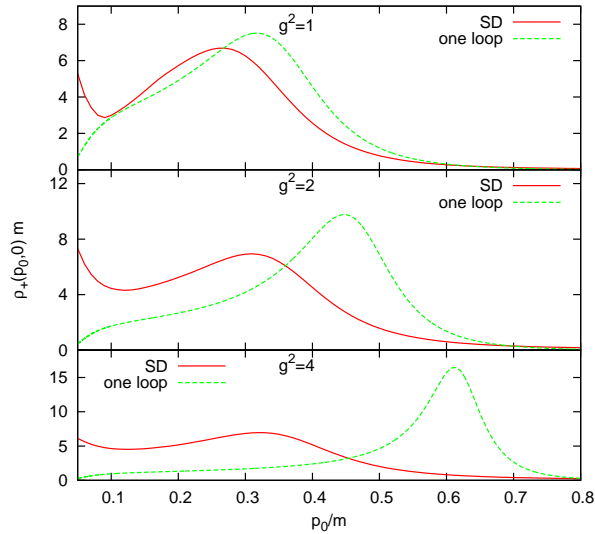}
\caption{The spectral functions $\rho_+(p_0,0) m$ for $T/m=1$ and
$g^2=1, 2, 4$ from top to bottom.}
\label{fig:peakg}
\end{figure}

\begin{figure}[ht]
\includegraphics[width=8cm]{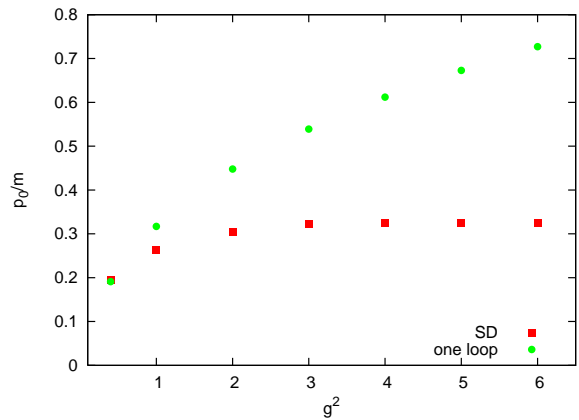}
\caption{The peak position of the normal quasifermion for $p=0$ and
$T/m=1$.}
\label{fig:peak2}
\end{figure}

In this subsection, we study the coupling dependence of the 
fermion spectrum, focusing on the peak structure.
In Fig.~\ref{fig:gspc}, we plot the fermion spectral function
$\rho_+$ for $T/m=1$ and $g^2=0.4, 2, 4$.
(For $T/m=1$ and $g^2=1$, see Fig.~\ref{fig:tspc}.)
The fermion spectrum for $g^2=0.4$ is closer to the free spectrum
than that for $g^2=1$ in Fig.~\ref{fig:tspc}, as it should be.
For $g^2>1$, all the peaks at the low momentum region become broader
and the height of the central peak falls rapidly.
These properties are consistent with our previous work in which
the fermion spectrum in a strong coupling gauge theory was investigated
with the SDE\cite{Harada:2007gg}:
Because the SDE contains effects of multiple scatterings with
bosons as mentioned in the previous subsection, the peak broadening
for larger coupling could be understood from this point of view.
The depression of the central peak is also understood from the
multiple scattering effects:
We plot the imaginary part of the fermion self-energy in Fig.~\ref{fig:sigg2}.
While the value at the origin is zero independently of the coupling
at one loop as mentioned previously, its absolute value from the
SDE gradually
increases as the coupling increases.
This means that higher order multiple scattering effects can bring
about the decay of the fermion with $p_0=p=0$ and enhances as the 
coupling increases.

Next we consider nonperturbative effects of the fermion spectrum
by comparing with the result of the one-loop calculation as shown
in Fig~\ref{fig:peakg}.
At $g^2=1$, the peak positions and the widths are close to each other.
At $g^2=2, 4$, the widths are broader than those at one loop,
which is due to the multiple scattering effects as mentioned above.
The peak positions are smaller than the one-loop results.
More specifically, we plot the coupling dependence of the peak
position at $p_0>0$ in Fig.~\ref{fig:peak2}.
We see that the peak position from the SDE is smaller than that
from the one loop and clearly saturates around $g^2=3$.
This behavior is consistent with that obtained in Ref.\cite{Harada:2007gg}.
In the HTL approximation, the width of the corresponding peak vanishes
and thus the peak position gives exactly the thermal mass which
is proportional to $T$ and the coupling.
Our results suggest that
a simple extrapolation of the results by the perturbation theory
to the strong coupling region is not valid for the thermal mass.

\section{Concluding remarks and outlook}
\label{sec:conc}

In this paper, we have investigated the fermion spectrum at finite
temperature ($T$) in 
a chiral U(1)${}_L\times$ U(1)${}_R$ invariant linear sigma
model employing coupled Schwinger-Dyson equations (SDE) for both
fermion and boson with the ladder approximation.
We formulate the SDE in the real time formalism and
thus the spectrum in both the timelike and spacelike regions
can be directly evaluated.
To reduce the computation task, only the imaginary parts of the
self-energies are evaluated with loop integrals, while the real
parts are evaluated from the imaginary parts with the dispersion
relation.

We have assumed the massless fermion and the massive boson and
focused on the $T$ and coupling dependences of the fermion spectrum.
The main reason is that it is pointed out from the one-loop evaluation
in Yukawa models that the three-peak structure
appears in the fermion spectrum at some $T$ and 
coupling\cite{Kitazawa:2006zi, Kitazawa:2007ep}.
The essential points of the existence of the three-peak structure
found in Ref.~\cite{Kitazawa:2006zi}
are the massiveness of the boson and the Landau damping at finite $T$.
One of the remaining issues is whether such a structure is stable
against higher order corrections, in particular, for the strong coupling
region.
Our present result supports the existence of the three-peak structure
even in the strong coupling region, although the strength of the central peak
rapidly suppresses owing to multiple scattering effects with bosons
as the coupling increases.
The quantitative difference between our result and the result from the
one-loop calculation is that
while the central peak at the origin is of the $\delta$-function shape
at one loop, it has a finite width which becomes broader as the coupling
and $T$ increase in the Schwinger-Dyson approach.

The present formalism can be readily applied to other models, such as gauge
theories in the ladder approximation with the fixed 
gauge.
Furthermore, because lattice QCD at high density is not realistic
in the present, the extension to finite density systems is also 
important and, in fact,  straightforward.

Because the soft bosonic modes are expected to arise dynamically 
near but above the critical temperature of the second order transition,
it would be also interesting to extend
this formalism in order to study effects of the phase transition
and temperature-dependent bosonic masses.
This should be possible in future study within the
presented dynamical framework.

\begin{acknowledgments}
This work is supported in part by the 21st Century COE Program at
Nagoya University and the Grant-in-Aid for 
Young Scientists No.18740140 (Y.N.) 
and the JSPS Grant-in-Aid for Scientific
Research (c) 20540262 (M.H.) by Monbu-Kagakusyo of Japan.
\end{acknowledgments}

\end{document}